\documentclass[xcolor=table]{elsarticle}

\usepackage{lineno,hyperref}
\modulolinenumbers[5]
\usepackage{subfigure}
\usepackage[table,xcdraw]{xcolor}
\usepackage{geometry}
\usepackage{amsthm,dsfont,amssymb,amsfonts}
\usepackage[fleqn]{amsmath}
\usepackage{graphicx}
\journal{Journal of XYZ}

\graphicspath{{figs/}}

\begin{document}

\begin{frontmatter}

\title{On stimulating fungi \emph{Pleurotus ostreatus} with Cortisol}

\author[1,2]{Mohammad Mahdi Dehshibi}
\author[1,3]{Alessandro Chiolerio}
\author[1,4]{Anna Nikolaidou}
\author[1]{Richard Mayne}
\author[5,6]{Antoni Gandia}
\author[2]{Mona Ashtari}
\author[1]{Andrew Adamatzky}

\address[1]{Unconventional Computing Laboratory, UWE, Bristol, UK}
\address[2]{Department of Computer Science, Universitat Oberta de Catalunya, Barcelona, Spain}
\address[3]{Center for Sustainable Future Technologies, Istituto Italiano di Tecnologia, Torino, Italy}
\address[4]{Department of Architecture, UWE, Bristol, UK}
\address[5]{Institute for Plant Molecular and Cell Biology, CSIC-UPV, Valencia, Spain}
\address[6]{Mogu S.r.l., Inarzo, Italy}

\begin{abstract}
Fungi cells are capable of sensing extracellular cues through reception, transduction and response systems which allow them to communicate with their host and adapt to their environment. They display effective regulatory protein expressions which enhance and regulate their response and adaptation to a variety of triggers such as stress, hormones, light, chemicals and host factors. In our recent studies, we have shown that \emph{Pleurotus} oyster fungi generate electrical potential impulses in the form of spike events as a result of their exposure to environmental, mechanical and chemical triggers, demonstrating that it is possible to discern the nature of stimuli from the fungi electrical responses. Harnessing the power of fungi sensing and intelligent capabilities, we explored the communication protocols of fungi as reporters of human chemical secretions such as hormones, addressing the question if fungi can sense human signals. We exposed \textit{Pleurotus} oyster fungi to cortisol, directly applied to a surface of a hemp shavings substrate colonised by fungi, and recorded the electrical activity of fungi. The response of fungi to cortisol was also supplementary studied through the application of X-ray to identify changes in the fungi tissue, where receiving cortisol by the substrate can inhibit the flow of calcium and, in turn, reduce its physiological changes. This study could pave the way for future research on adaptive fungal wearables capable for detecting physiological states of humans and biosensors made of living fungi.
\end{abstract}


\end{frontmatter}


\section{Introduction}
All living organisms have evolved elaborate communication processes and mechanisms to sense, respond, and adapt to the surrounding environment in order to survive. These processes take place through reception, transduction, and response systems, which enable them to sense and adapt to their surroundings in response to a variety of cues such as nutrients, light, gases, stress, and host factors. Any form of communication requires the existence of three essential elements: a sender, a message, and a receiver. The process begins with a sender releasing a message and ends with a receiver understanding the message~\cite{cottier2012communication}. Fungi are composed of eukaryotic cells that report, react, and adapt to external stimuli primarily through signal transduction pathways~\cite{alonso2009fungi}. They have extracellular and intracellular sensing mechanisms, as well as protein receptors that enable them to detect and respond to a variety of signals. \textit{Pleurotus ostreatus}, a basidiomycete fungi, has effective regulatory protein expression that enhances its adaptation to stress triggers~\cite{hou2019expression}.

In our previous studies, we reported that the oyster fungi \textit{Pleurotus djamor} exhibit trains of electrical potential spikes similar to action potential spikes~\cite{adamatzky2018spiking,beasley2020capacitive,beasley2020fungal,adamatzky2021electrical}. Our initial assumption was that spike trains might reflect increasing mycelium propagation in the substrate, nutrient and metabolite transport, and communication processes within the mycelium network. We investigated the information-theoretic complexity of fungal electrical activity~\cite{adamatzky2018towards,adamatzky2020fungal,dehshibi2021electrical} to pave the way for additional investigation into sensorial fusion and fungi decision making~\cite{adamatzky2019fungal,adamatzky2020boolean,goles2020computational}. Later, in a series of laboratory experiments~\cite{adamatzky2021fungal,adamatzky2021reactive}, we demonstrated that fungal electrical activity patterns, specifically mycelium bound hemp composite, changes in response to stimuli such as light, mechanical stretching, and attractants and repellents. Our findings demonstrated that fungi are a promising candidate for producing sustainable textiles for use as eco-friendly bio-wearables.

We present an illustrative scoping study in which we investigate the short and long-term dynamics in mycelium of the oyster fungi \emph{Pleurotus ostreatus} in response to stimulation with cortisol. The purpose of this study is to enthuse the scientific community to address the issue of fungi being able to sense animal hormones. The human body's adrenal glands release hormones such as cortisol and adrenaline. Cortisol levels in various bodily fluids can range from 4~pM to 70~pM~\cite{jang2018electronic,kaushik2014recent}. Sweat cortisol levels have a strong correlation with salivary cortisol concentrations~\cite{russell2014detection}, and the optimal cortisol level ranges from 0.02 to 0.5~M~\cite{jang2018electronic,kaushik2014recent}. Monitoring cortisol levels in bodily fluids, which can be altered by chronic stress and disease, is critical for maintaining healthy physiological conditions. For this study, not only are the electrical activities investigated, but the substrate is also irradiated with the X-ray spectrum from multiple angles to produce cross-sectional images of the substrate. This multimodal approach enables us to identify and track the dynamics of changes in the tissue of the mycelium anatomy.

We demonstrated that fungi's electrical responses and reconstructed computed tomography images can be used to detect the presence of stimuli. The findings could lead to the development of biosensing patches for use in organic electronics and bio-electronics, especially with living substrates, which offer a great opportunity to integrate natural systems' parallel sensing and information processing capabilities into future and emerging wearables.

The rest of this paper is structured as follows. Section~\ref{sec:2} presents the experimental setup and details of the analysis. Experimental results are discussed in Sect.~\ref{sec:3}. Finally, the discussion is given in Sect.~\ref{sec:4}.

\section{Methods} \label{sec:2}
\subsection{Experimental setup}

\begin{figure}[!htbp]
    \centering
    \subfigure[]{\includegraphics[width=0.45\textwidth]{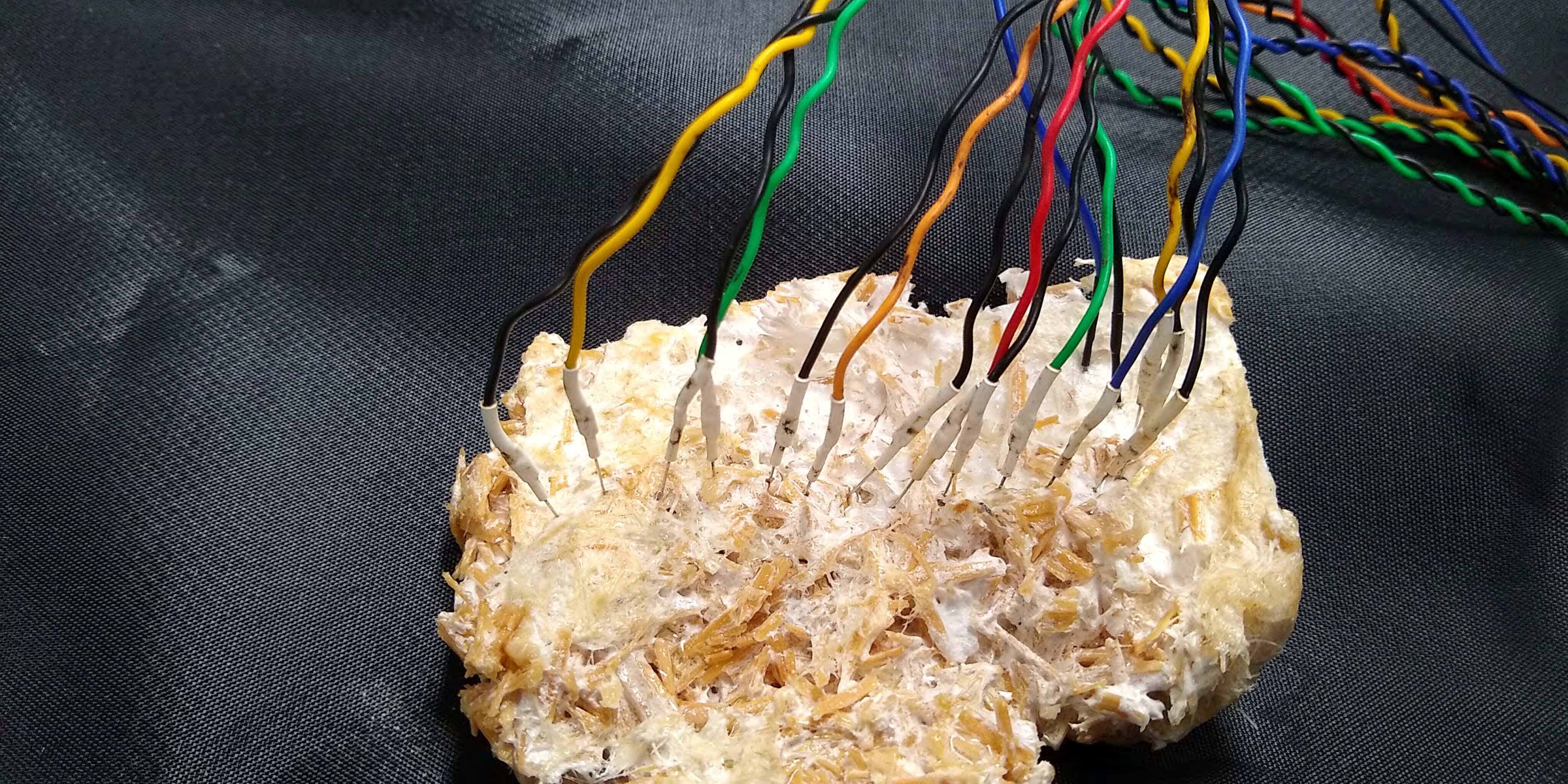}\label{fig:electrodes}}
    \subfigure[]{\includegraphics[width=0.45\textwidth]{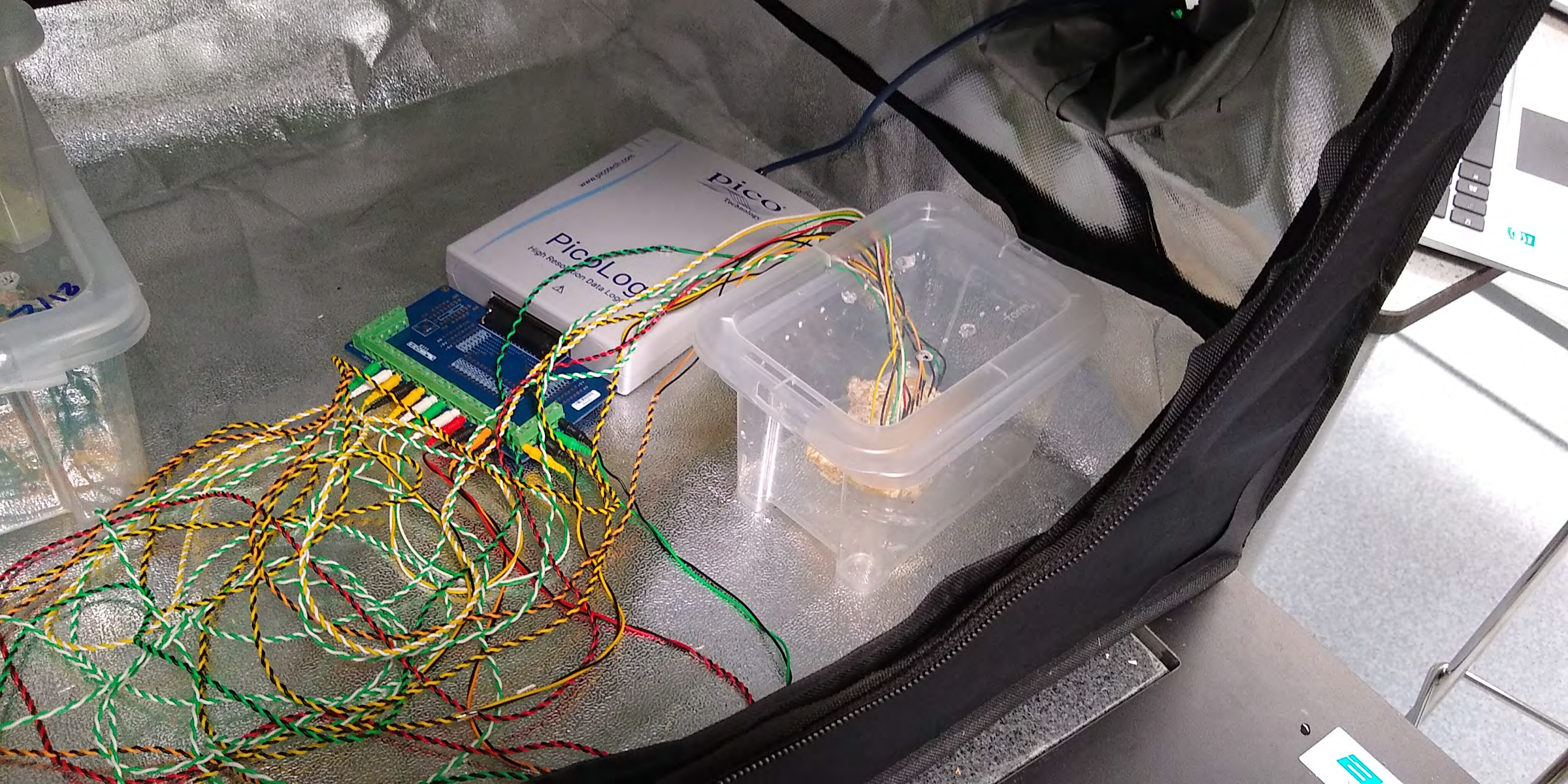}\label{fig:tent}}        
    \caption{ Experimental setup.
    (a)~Exemplar locations of electrodes.
    (b)~Electrode pairs and logging setup in the tent.
  }
    \label{fig:setup}
\end{figure}

A commercial strain of the fungus \textit{Pleurotus ostreatus} (Mogu's collection code 21-18), preselected for showing a superior fitness growing on different lignocellulosic substrates, was cultured on sterilised hemp shives contained in plastic boxes c.~$35 \times 20~\mathrm{cm}^2$ in darkness at ambient room temperature c.~$22^\circ$C. Particles of substrate well colonised by the fungus were spread on rectangular fragments, c. $12 \times 12~\mathrm{cm}^2$, of moisturised nonwoven hemp fibre mats. When the mats were properly colonised, as visually confirmed by white and homogeneous mycelial growth on the surface, these were used in the experiments. The humidity of the hemp mats ranged from 70\% to 80\% (MerlinLaser Protimeter, UK). The experiments were carried out in a room with an ambient temperature of c.~$21^\circ$C in the absence of light. Figure~\ref{fig:setup} shows examples of the experimental setups.

Electrical activity of the colonised hemp mats was recorded using pairs of iridium-coated stainless steel sub-dermal needle electrodes (Spes Medica S.r.l., Italy) with twisted cables, via a high-resolution data logger with a 24-bit A/D converter, galvanic isolation, and software-selectable sample rates (Pico Technology, UK). The electrodes were placed in a straight line with a distance of 1-2~cm. To keep the electrodes stable, we put a polyurethane pad underneath the fabric. As a result, the electrode pairs were inserted through the fabric and onto the polyurethane pad, as shown in Fig.~\ref{fig:electrodes}.

In each trial, we recorded the electrical activity of seven electrode pairs simultaneously. Each pair of electrodes, referred to as a Channel (Ch), reported a difference in the electrical potential between the electrodes. The electrical activity was recorded at a rate of one sample per second (1~Hz), with logging times ranging from 60.04 to 93.45 hours. Throughout the recording, the logger took as many samples as it could (typically up to 600 per second) and saved the average value. We set the acquisition voltage range to 156~mV with an offset accuracy of 9~$\mu$V at 1~Hz to maintain a gain error of 0.1\%. Each electrode pair was considered independently with a noise-free resolution of 17 bits and conversion time of 60~ms. Figure~\ref{fig:tent} shows one the recording setup inside a light-proof growing tent.

For stimulation, we used hydro-cortisone (Solu-Cortef trademark, 4~mL Act-O-Vial, Pfizer, Athens, Greece). We then applied 2~mL of the resulting solution to the surface of the colonised substrate in the loci surrounding Ch4 and Ch5. 

The following was the rationale behind the dosage selection: human patients weighing 80~kg are typically given 20~mg of synthetic cortisol per day for a variety of diseases, such as maintenance therapy for patients with hypopituitarism. As a result, 20~mg divided by 80~kg yields 0.25~mg/kg. Hemp mats colonised with fungi weighed around 100-200~g each, so a 2~mL dose of 250~$\mu$g was needed.

\subsection{Electrical Activity Analysis}

\begin{figure}[!htbp]
    \centering
    \includegraphics[width=0.8\linewidth]{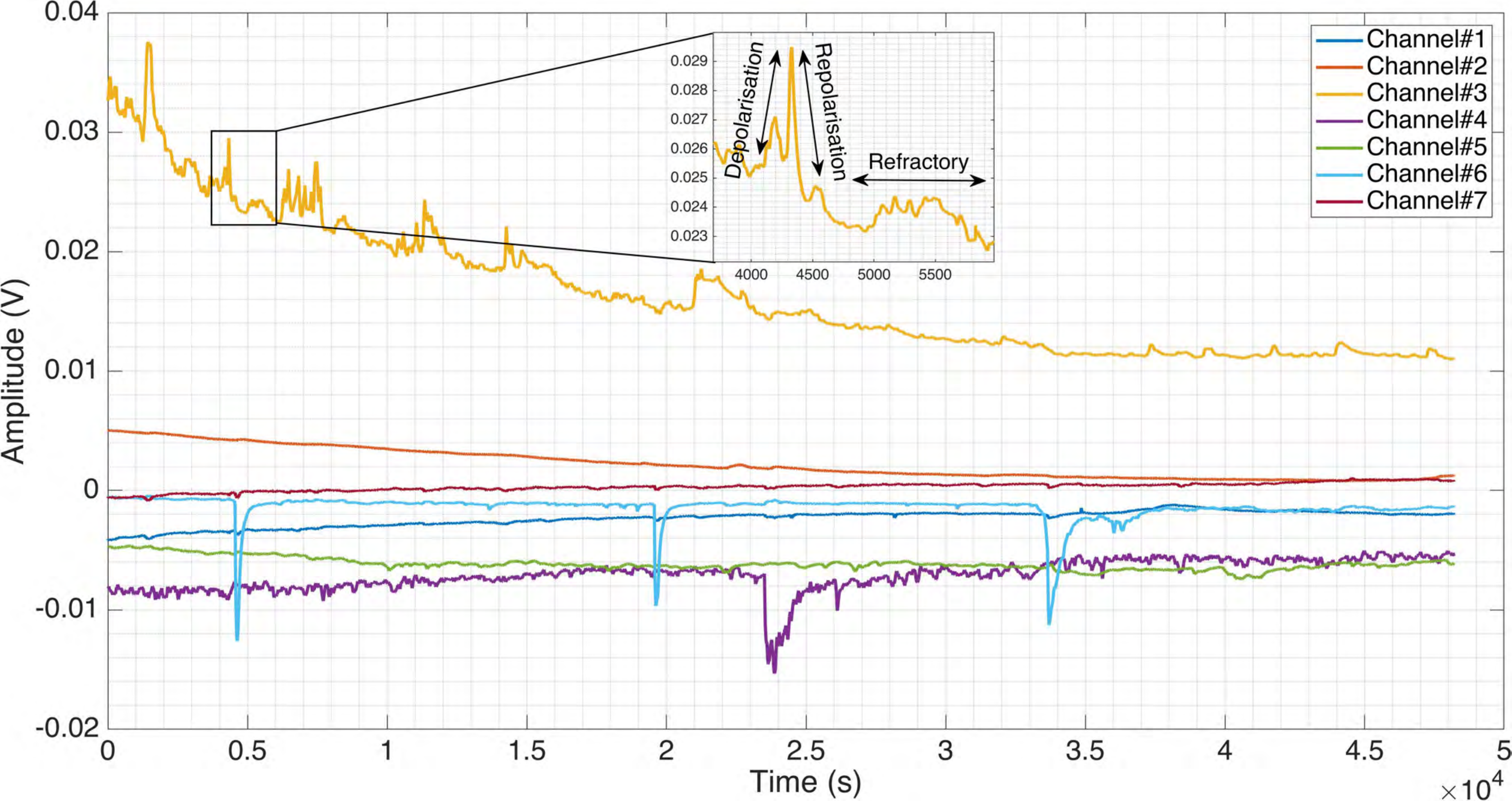}
    \caption{An example of electrical activity observed in seven channels of colonised hemp mats with fungi over 13.3 hours. The inserts are zoomed in on one channel to display a spike event with `depolarisation', `repolarisation', and `refractory' cycles. This spike has a duration of 2258 seconds and a refractory time of 1426 seconds. The rates of depolarisation and repolarisation are 54.85 and 45.75$\mu\mathrm{V/s}$, respectively.}
    \label{fig:spike}
\end{figure}

Extracellular signals which surpassed a certain amplitude threshold with depolarisation, repolarisation, and refractory cycles are referred to as spike events (see Fig.~\ref{fig:spike}). Spike events represent the physiological and morphological processes of mycelium in the colonised hemp mat. We proposed a novel method for identifying spiking events, including three main stages as (1) splitting signal into chunks, (2) smoothing the chunk by mapping the constant amplitude to an instantaneous amplitude, and (3) detecting spike events.

Assume $\mathcal{X} = \{(t_{i}, x_{i})\}_{i=1}^{\mathbb{C}}$ is a recording set of $\mathbb{C}$ channels with the entire length of $T$ seconds and samplings rate of $f_{s}$~Hz, where $x$ defines the signal's sample value at time $t,~ 1 \leq t \leq T$. Our objective is to detect the set of spike events $\mathcal{S} = \{s_{1}, s_{2}, \cdots, s_{\eta}\}$, where $\eta << T$. We segmented the signal $\mathcal{X}$ using the idea of the variable size sliding window to analyse its dynamics before and after cortisol application in the same intervals. We split the signal into two segments, containing the electrical activity of the channel before and after cortisol application. Each segment was then subdivided into chunks of $m=\{1,2,4,8,16\}$ hours. Note that the chunk $m=16$ in all 16 experiments does not have the same length that resulted in 12 to 16 hours of electrical activity recording.

The electrical activity of the colonised hemp mats with fungi displays diffraction patterns. The presence of these diffraction patterns from multiple slits can cause determining spike events to be distorted. The envelope of an oscillating signal can expand the concept of constant amplitude to instantaneous amplitude and, therefore, bypass multiple slits with a single slit diffraction pattern to outline significant extremes, i.e., spike events~\cite{dehshibi2021electrical}.

To obtain the signal envelope ($\xi$), we used the discrete Fourier Transform, as implemented in the Hilbert transform, to detect the analytical signal. Then, inspired by Marple et al.~\cite{marple1999computing}, we set the negative frequency in half of each spectral period to zero, resulting in a periodic one-sided spectrum. More formally, using the sampling theorem~\cite{shannon1949communication}, we convert the input chunk\footnote{Note that here we intentionally drop the $m$ superscript to simplify mathematical notations.}, $\mathcal{X}$, into a sequence of values with the sample period of $\tau \triangleq w\frac{T}{f_{s}}$ (see Eq.~\ref{eq:01}). 

\begin{equation}
    \label{eq:01}
    L[n] \triangleq \tau \cdot x(n\tau),~ 0 \leq n \leq N-1
\end{equation}
where $N$ is an even number corresponding to the number of discrete-time analytical signal points. To obtain the $N$-point one-sided discrete-time analytic signal using Hilbert transform~\cite{marple1999computing}, we need to calculate the discrete-time Fourier transform of $L[n]$, with sampling at $\tau$ intervals to prevent aliasing (see Eq.~\ref{eq:02}). We take the second numerical signal derivation ($L = \frac{\partial^2 \mathcal{X}}{4\partial t^2}$) to highlight effective signal peaks and neutralise diffraction patterns.

\begin{equation}
    \label{eq:02}
    F(\omega) = \tau \sum_{n=0}^{N-1}L[n]e^{-i2\pi \omega \tau n}
\end{equation}
where $|\omega| \leq \frac{1}{2\tau}$~Hz. To obtain a periodic one-sided spectrum ($Z[k]$), we set the negative frequency in half of each spectral period to zero and calculate the spectrum using Eq.~\ref{eq:03}.

\begin{equation}
    \label{eq:03}
    Z[k] = \begin{cases}
            F[0], & \text{ for } k = 0\\ 
            2F[k], & \text{ for } 1 \leq k \leq\frac{N}{2}-1 \\ 
            F[\frac{N}{2}], & \text{ for } k = \frac{N}{2}\\ 
            0, & \text{ for } \frac{N}{2}+1 \leq k \leq N-1.
    \end{cases}
\end{equation}

\begin{figure}[!htb]
    \centering
    \includegraphics[width=0.8\linewidth]{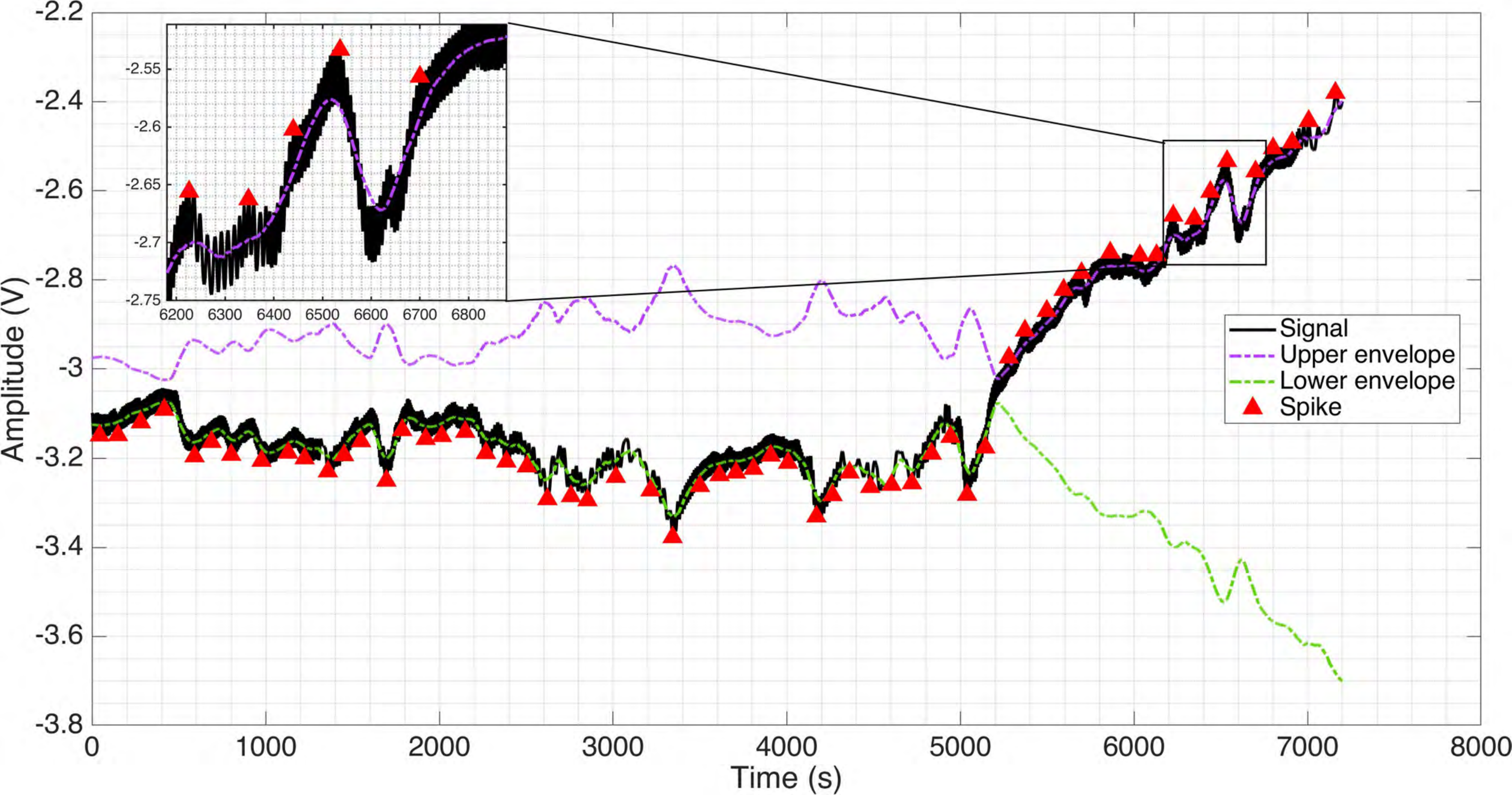}
    \caption{Construct the intermediate representation of an input signal using Eq.~\ref{eq:06}, where the upper and lower envelopes are shown in violet and green dash lines, respectively. We zoomed in on one chunk electrical activity to highlight the identified spike events.} \label{fig:spikeevent}
\end{figure}

To obtain the envelope of the original signal $x(t)$, we need to calculate the inverse discrete-time Fourier transform of $F(\omega)$ and $Z[k]$ (see Eq.~\ref{eq:04}). 

\begin{equation}
    \label{eq:04}
    x_{a}[n] \triangleq \mathcal{F}^{-1}\left [ F(\omega) \right ],~~~~~~ z_{a}[n] \triangleq \mathcal{F}^{-1}\left [ Z[k] \right ].
\end{equation}
where $\mathcal{F}^{-1}[\cdot]$ takes the inverse of the Fourier transform, $x_{a}$ and $z_{a}$ are the analytical signals, and $z_{a}[n] = \frac{1}{N\tau}\sum_{k=0}^{N-1}Z[k]e^{\frac{i2\pi kn}{N}}$. By taking the root-mean-square value of the analytical signals as in Eq.~\ref{eq:05}, we can calculate $\xi$ for the signal $\mathcal{X}$.

\begin{align}
    \label{eq:05}
    e_{a}[n] = x_{a}[n] + jz_{a}[n],\nonumber \\
    \xi[n] = \sqrt{e_{a}[n] \times \bar{e}_{a}[n]}.
\end{align}
where $\bar{e}_{a}$ is the the complex conjugate of $e_{a}$, and $j$ refers to the imaginary part of the analytical signal. We construct an intermediate representation ($\tilde{x}$) of the input signal using the upper and lower envelopes to identify spike events. All amplitude values greater than or equal to the upper envelope and less than or equal to the lower envelope are replaced with the upper and lower envelope values, respectively, while all other amplitude values are preserved. Then, as in Eq.~\ref{eq:06}, we calculate the absolute differences between the input signal ($x$) and this intermediate representation ($\tilde{x}$).

\begin{align}
    \label{eq:06}
    \tilde{x}[n] = \begin{cases}
            \xi_{l}[n], & x[n] \leq \xi_{l}[n]\\ 
            \xi_{u}[n], & x[n] \geq \xi_{u}[n] \\ 
            x[n], & \text{otherwise}.
    \end{cases} \nonumber \\
    \Phi[n] = \left | x[n]-\tilde{x}[n] \right |.
\end{align}
where $\xi_{l}$ and $\xi_{u}$ are the lower and upper envelopes, respectively. We locate all local maxima (peaks) of the $\Phi[n]$ where the minimum peak prominence is $\gamma$, and the minimum distance between is $w$. Here, $\gamma$ is the 99\% of confidence interval calculating using Eq.~\ref{eq:07}.

\begin{equation}
    \label{eq:07}
    \gamma = {\bar{\Phi}[n]}+z^{*}\frac{\sigma}{\sqrt{N}}.
\end{equation}
where $z^{*}=2.576$~\cite{dekking2005modern}, and $\bar{\Phi}[n]$ and $\sigma$ are the mean and standard deviation of $\Phi[n]$. Figure~\ref{fig:spikeevent} shows an example of the proposed method's results.

\subsection{CT Images Analysis}

\begin{figure}[!htbp]
    \centering
    \subfigure[]{\includegraphics[width=0.70\textwidth]{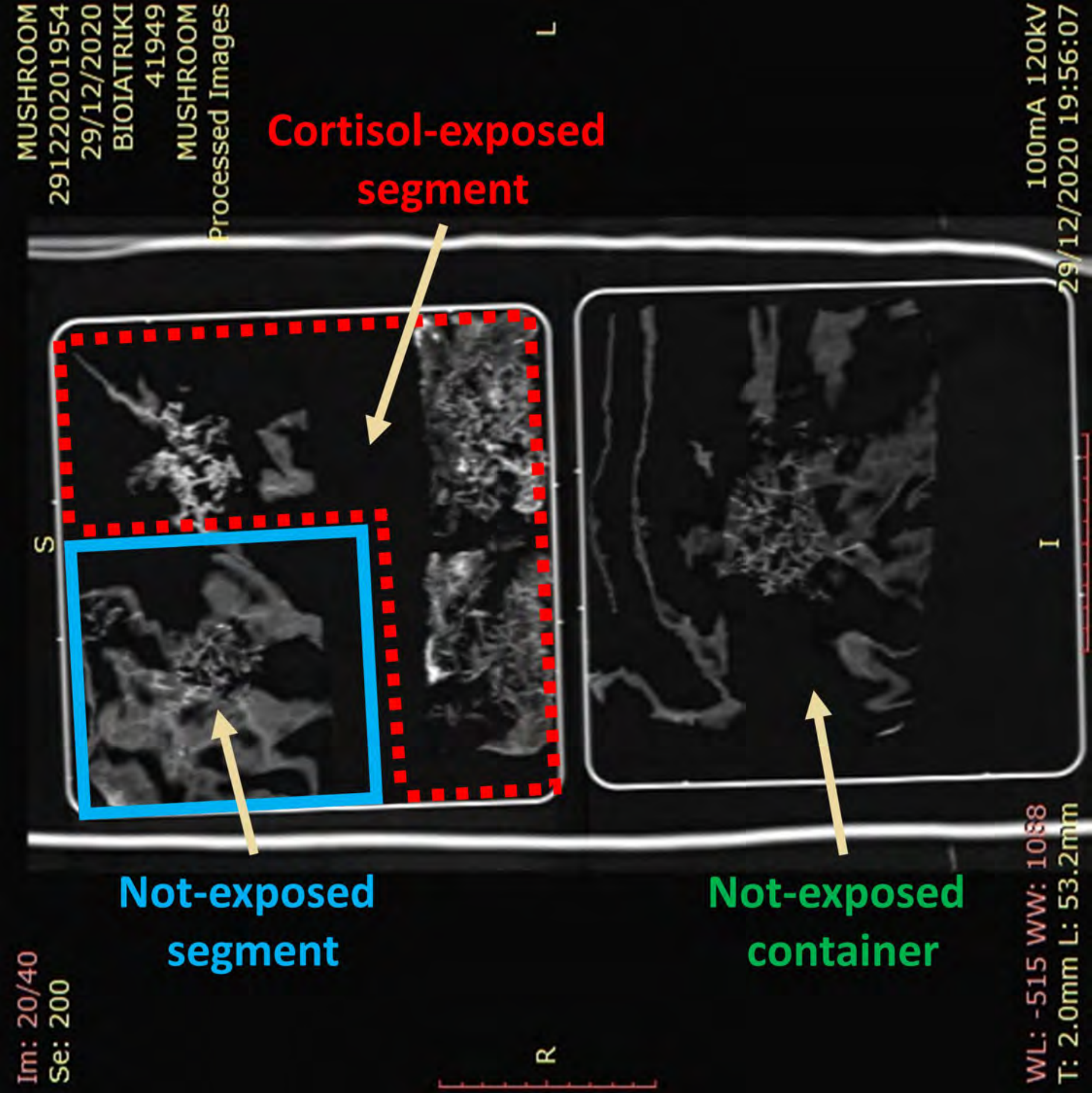}\label{fig:exposemark}}
    \subfigure[]{\includegraphics[width=0.38\textwidth]{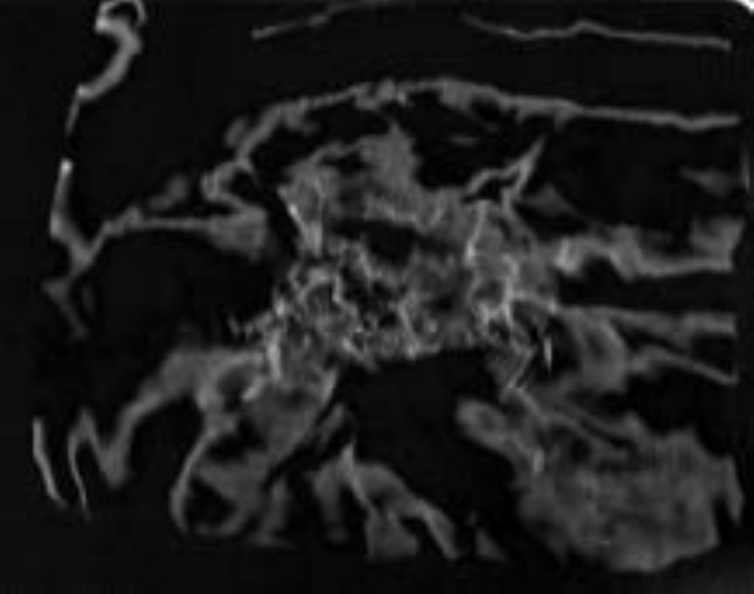}\label{fig:noexposure}}
    \subfigure[]{\includegraphics[width=0.35\textwidth]{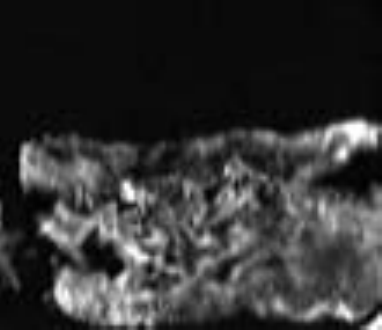}\label{fig:ctcortisol}}
    \caption{(a) A CT image of two containers, with the substrate in the left container receiving cortisol two hours before being irradiated with the X-ray spectrum, and the substrate in the right container receiving no stimulus. (b) The substrate segment in the left container that is not directly exposed to stimuli. (c) The cortisol-exposed segment of the substrate in the left container.}
    \label{fig:CT}
\end{figure}

We prepared two containers to obtain a better understanding of the effect of stimuli on the fungal substrate. One container received no stimuli, while the other received cortisol two hours before being irradiated with an X-Ray spectrum (see Fig.~\ref{fig:exposemark}). We used two different containers since the fungal colony retains its integrity through the flow of cytoplasm in the mycelium network, where calcium waves~\cite{aramburu2004calcineurin} and associated electrical potential waves change the propagation coordinate of this flow. Therefore, we were able to analyse the stimuli spread across the flow of cytoplasm by comparing the CT image of the container that had not received any stimulus with the container that had only a portion of it remained unexposed. We also segmented the container's image that was exposed by the stimulus into four segments (see Figs.~\ref{fig:noexposure}--\ref{fig:ctcortisol}) to investigate the influence of each stimulus separately.

\begin{figure}[!htbp]
    \centering
    \includegraphics[width=\textwidth]{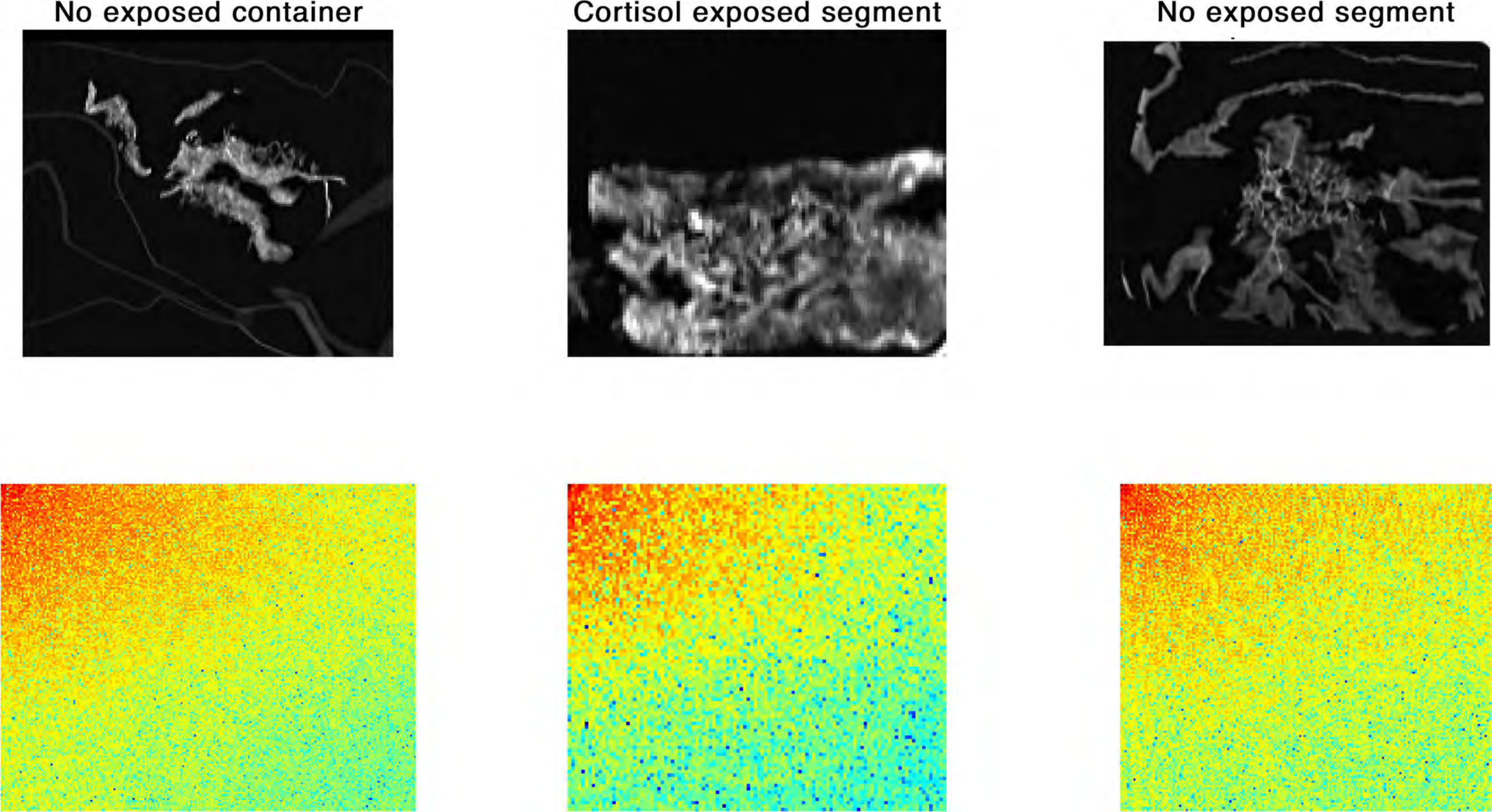}
    \caption{The results of applying DCT to CT images, where DCT expresses a finite sequence of data points as a sum of cosine functions oscillating at different frequencies. (a) On a non-exposed container, higher energies (red hue) are concentrated in the top left corner, while lower energies (green and blue hues) are concentrated in the bottom right corner. (b) The cortisol-exposed segment has a lower concentration of high energy in the top left corner and a higher dispersion of low energy (blue hue). (c) The no-exposed segment has a moderate energy distribution than the others, where the higher-energy spectral sub-bands are scattered rather than being concentrated in a few DCT coefficients.}
    \label{fig:DCT}
\end{figure}

When working with images in the spatial domain, we deal with changes in pixel values with respect to the scene. In the frequency domain, however, we are interested in the rate at which the pixel values in the spatial domain change, which provides us with a better understanding of the underlying distribution of changes and complexities. The discrete cosine transform (DCT) was used in this study (see Eq.\ref{eq:08}) to transform the input CT image $I_{R \times C}$ from the spatial domain to the frequency domain, allowing us to divide the image into spectral sub-bands of varying significance. This transform can also concentrate the majority of the image's visually important details in just a few DCT coefficients. This property enables us to uncover changes in the substrate that are not apparent throughout visual inspection.

\begin{align}
    \label{eq:08}
    DCT_{pq} = \alpha_{p} \alpha_{q} \sum_{r=0}^{R-1} \sum_{c=0}^{C-1} I_{rc} \cos\frac{\pi\left ( 2r+1 \right )p}{2R} \cos\frac{\pi\left ( 2c+1 \right )q}{2C}, \nonumber \\ 
    \alpha_{p} = \left\{\begin{matrix}
    1/\sqrt{R} & p=0\\ 
    \sqrt{2/R} & 1 \leq p \leq R-1
    \end{matrix}\right. ~~ \alpha_{q} = \left\{\begin{matrix}
    1/\sqrt{C} & q=0\\ 
    \sqrt{2/C} & 1 \leq q \leq C-1
    \end{matrix}\right..
\end{align}
where $DCT_{pq}$ represent the DCT coefficients of $I_{R \times C}$, $R$ is the number of rows, and $C$ is the number of columns. Figure~\ref{fig:DCT} shows three examples of the results obtained by applying DCT to CT images.

\subsection{Statistical and Information Theory Metrics}

In statistics~\cite{fan1998local} and information theory~\cite{petz2007quantum}, a concept of entropy is essential. In statistics, entropy refers to the inference functional for an updating process, and in information theory, it specifies the shortest attainable encoding scheme. Recent advances in complex dynamical systems, on the other hand, have necessitated an extension of the entropy theory beyond the conventional Shannon-Gibbs entropy ($H$)~\cite{shannon1948mathematical}. In this study, we used the R\'{e}nyi ($R_{q}$)~\cite{renyi1961measures} and Tsallis ($T{q}$)~\cite{tsallis1988possible} additive entropy concepts, which are generalisations of the classical Shannon entropy. Regardless of the generalisation, these two entropy measurements are used in conjunction with the Principle of maximum entropy, with entropy's main application being in statistical estimation theory. Shannon, Tsallis, and R\'{e}nyi entropy measurements are expressed as in Eq.~\ref{eq:09}.

\begin{align}
    \label{eq:09}
    H(\mathcal{S})=-\sum_{i}p(s_{i})\log p(s_{i}), \nonumber\\
    T_{q}(p_{i}) =\frac{k}{q-1} \left( 1 - \sum_{i} p_{i}^{q} \right), \\
    R_{q}(\mathcal{S})=\frac {1}{1-q}\log \left (\sum _{i}p_{i}^{q} \right ). \nonumber
\end{align}

Here, $\mathcal{S}$ is a discrete random variable that represents spike events, with potential outcomes in the set $\mathcal {S}=\{s_{1}, s_{2},\cdots,x_{\eta}\}$ and corresponding probabilities $p_{i} \doteq \Pr(\mathcal{S}=s_{i})$ for $i=1,\cdots ,\eta$. $p_{i}$ is a discrete set of probabilities with the condition $\sum_{i} p_{i}=1$, and $q$ is the \textit{entropic-index} or R\'{e}nyi entropy order with $q \geq 0$ and $q \neq 1$, which in our experiments was set to 2. We take the logarithm to be in base 2, since we interpreted the entire recording duration $T$ with bits, with `1s' indicating the availability of spike events and `0s' otherwise.

To determine spike diversity across all channels, we represent recordings by a binary matrix with a row for each channel, where '1s' indicate spike events and `0s' otherwise. We calculated the Lempel--Ziv complexity ($LZc$) over channels using the Kolmogorov complexity algorithm~\cite{kaspar1987easily} to capture both temporal and spatial diversity. We then concatenated the rows of this binary matrix to form a single binary string and normalised $LZc$ by dividing the raw value by the randomly shuffled value obtained for the same binary input sequence to obtain PCIpK. Since the value of PCIpK for a fixed-length binary sequence is maximum when it is totally random, the normalised values reflect the degree of signal diversity on a scale of 0 to 1.

Other metrics that were used to quantify the complexity of spike events are Simpson's diversity, Space-filling, and Expressiveness. Simpson's diversity is calculated as $\Gamma=\sum_{i}(p(s_{i}))^{2}$. For $H<3$, it correlates linearly with Shannon entropy, and the relationship becomes logarithmic for higher values of $H$. The value of $\Gamma$ varies from 0 to 1, with 1 representing infinite diversity and 0 representing no diversity. Space-filling ($\Delta$) is the ratio of non-zero entries in the binary representation of $\mathcal{X}$ to the total duration of the recording $T$. Expressiveness ($\Upsilon$) is calculated as the Shannon entropy $H$ divided by space-filling ratio $\Delta$, which represents the economy of diversity.

\section{Results} \label{sec:3}

Table~\ref{tbl:statistic} presents a qualitative description of electrical activity for one of the trials, where all complexity metrics can be compared for the condition in which the substrate was exposed 16 hours before cortisol exposure and 1 hour after cortisol exposure.

\begin{table}[!t]
\centering
\caption{Qualitative description of electrical activity (1) 16 hours prior to cortisol exposure and (2) 1 hour after cortisol exposure.}
\label{tbl:statistic}
\resizebox{\textwidth}{!}{%
\begin{tabular}{lcccccccccccc}
\hline
Ch\# & Spike\# & Length (s) & Amplitude (V) & $H(\mathrm{signal})$ & $H(\mathrm{spike})$ & $\Gamma$ & $\Delta$ & $\Upsilon$ & Kolmogorov & PCIpK & $T_{q}$ & $R_{q}$ \\ \hline
\multicolumn{13}{l}{\cellcolor[HTML]{C0C0C0}16 hours prior to the trigger event} \\ \hline
1 & 455 & 2.91 & -1.20 & -3.40 & 242.2 & 0.99 & 0.0078 & $3.07 \times 10^{4}$ & 0.055 & $3.86 \times 10^{-03}$ & -3.230 & -0.62 \\
2 & 449 & 2.76 & -0.32 & -3.72 & 239.6 & 0.99 & 0.0077 & $3.07 \times 10^{4}$ & 0.052 & $3.64 \times 10^{-03}$ & -2.498 & -0.29 \\
3 & 457 & 2.72 & -0.02 & -3.88 & 243.1 & 0.99 & 0.0079 & $3.06 \times 10^{4}$ & 0.053 & $3.66 \times 10^{-03}$ & -0.001 & 12.52 \\
4 & 449 & 3.40 & -3.58 & -3.74 & 239.6 & 0.99 & 0.0077 & $3.07 \times 10^{4}$ & 0.052 & $3.73 \times 10^{-03}$ & -46.884 & -3.70 \\
5 & 460 & 2.23 & 1.18 & -3.43 & 244.5 & 0.99 & 0.0079 & $3.06 \times 10^{4}$ & 0.055 & $3.64 \times 10^{-03}$ & 0.002 & 12.27 \\
6 & 446 & 3.19 & -1.09 & -3.26 & 238.3 & 0.99 & 0.0077 & $3.08 \times 10^{4}$ & 0.052 & $3.69 \times 10^{-03}$ & -6.544 & -1.49 \\
7 & 464 & 2.73 & 0.26 & -3.58 & 246.2 & 0.99 & 0.0080 & $3.06 \times 10^{4}$ & 0.054 & $3.59 \times 10^{-03}$ & 0.102 & 4.30 \\ \hline
\multicolumn{13}{l}{\cellcolor[HTML]{C0C0C0}1 hour after the trigger event} \\ \hline
1 & 29 & 2.38 & -1.08 & -5.72 & 20.60 & 0.96 & 0.0080 & $2.55 \times 10^{3}$ & 0.062 & 0.048 & -3.68 & -0.78 \\
2 & 28 & 2.97 & -0.80 & -5.49 & 20.01 & 0.96 & 0.0077 & $2.57 \times 10^{3}$ & 0.062 & 0.047 & -1.93 & 0.04 \\
3 & 30 & 2.52 & -0.14 & -5.38 & 21.18 & 0.96 & 0.0083 & $2.54 \times 10^{3}$ & 0.065 & 0.047 & -0.12 & 4.43 \\
4 & 27 & 3.15 & -6.15 & -5.56 & 19.41 & 0.96 & 0.0075 & $2.58 \times 10^{3}$ & 0.065 & 0.051 & -47.67 & -3.71 \\
5 & 28 & 1.41 & -0.22 & -5.73 & 20.01 & 0.96 & 0.0077 & $2.57 \times 10^{3}$ & 0.062 & 0.045 & -0.10 & 4.64 \\
6 & 32 & 2.42 & -2.43 & -5.70 & 22.34 & 0.96 & 0.0088 & $2.51 \times 10^{3}$ & 0.075 & 0.049 & -8.47 & -1.79 \\
7 & 29 & 2.03 & 0.25 & -5.51 & 20.60 & 0.96 & 0.0080 & $2.55 \times 10^{3}$ & 0.068 & 0.050 & 0.23 & 1.83 \\ \hline
\end{tabular}%
}
\end{table}

\begin{figure}[!h]
    \centering
    \subfigure[]{\includegraphics[width=0.4\textwidth]{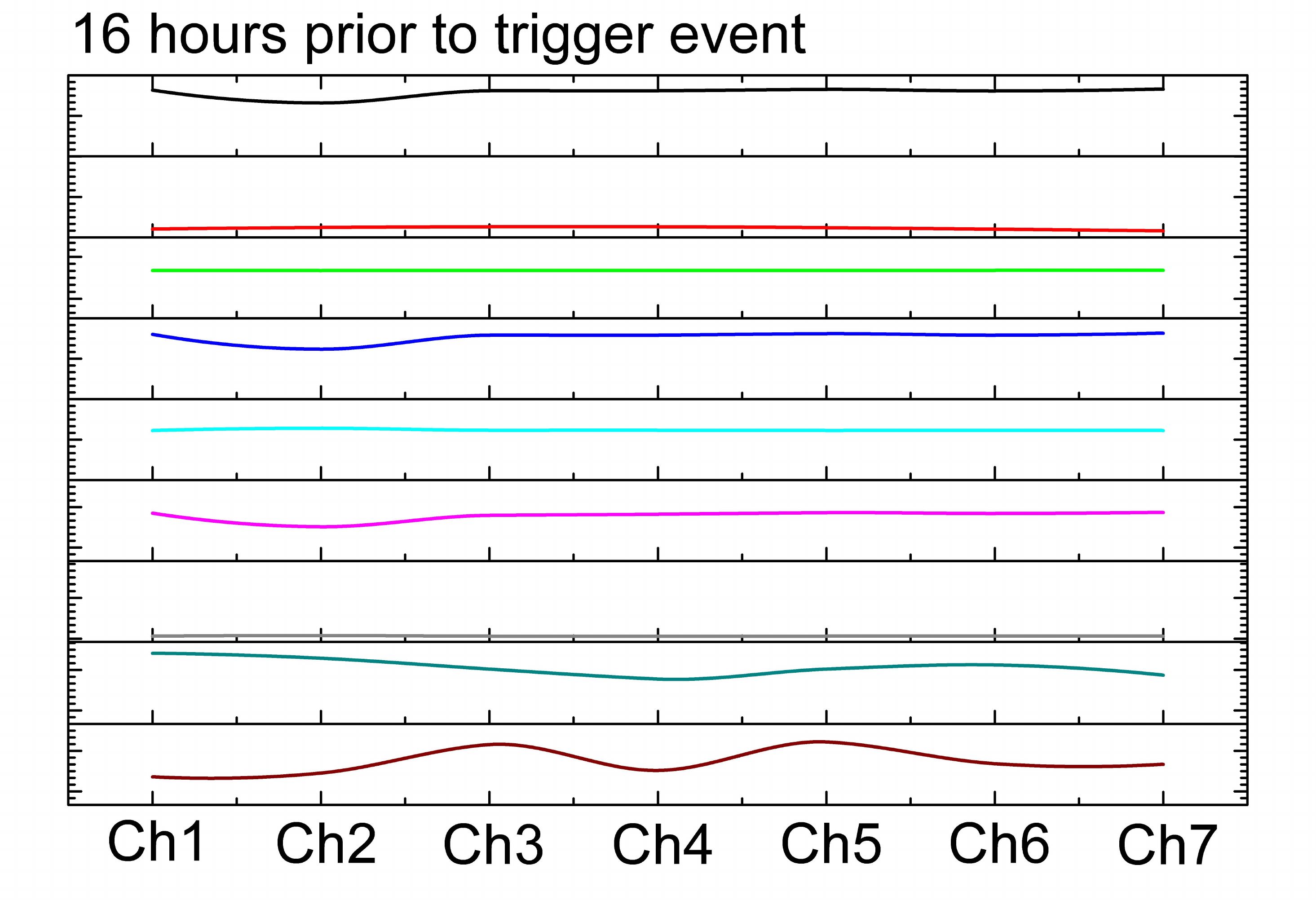}}
    \subfigure[]{\includegraphics[width=0.4\textwidth]{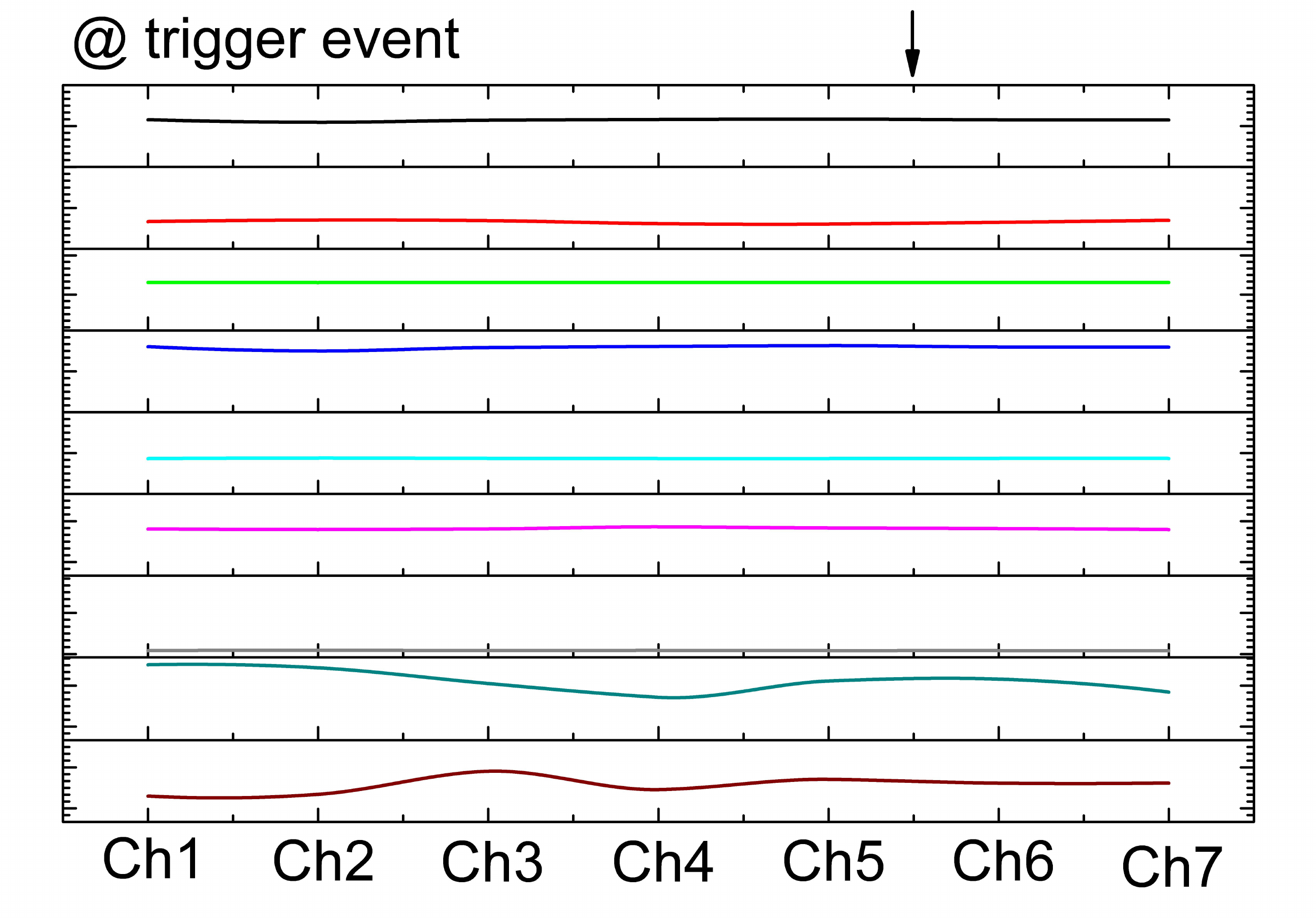}}
    \subfigure[]{\includegraphics[width=0.4\textwidth]{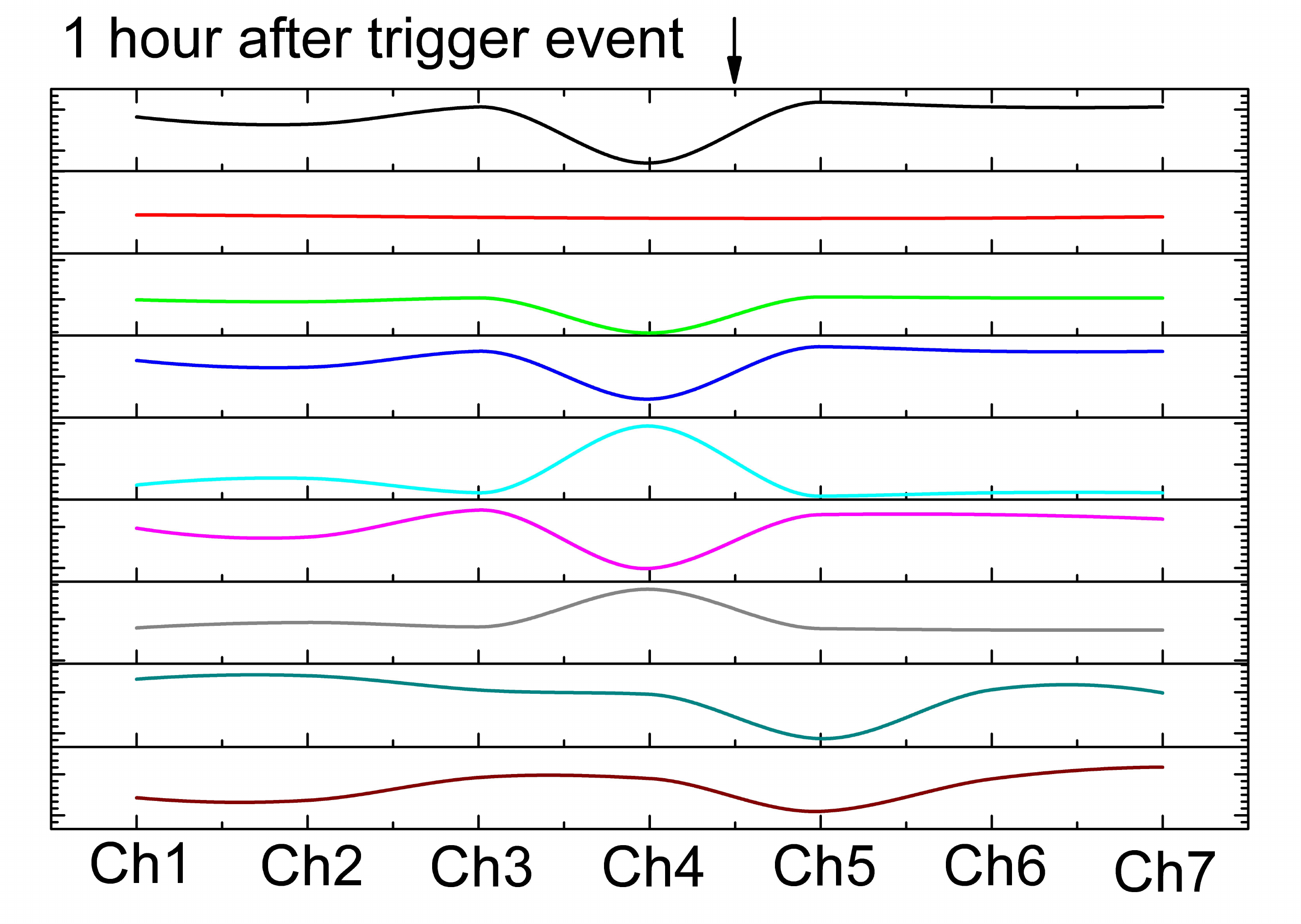}}
    \subfigure[]{\includegraphics[width=0.32\textwidth]{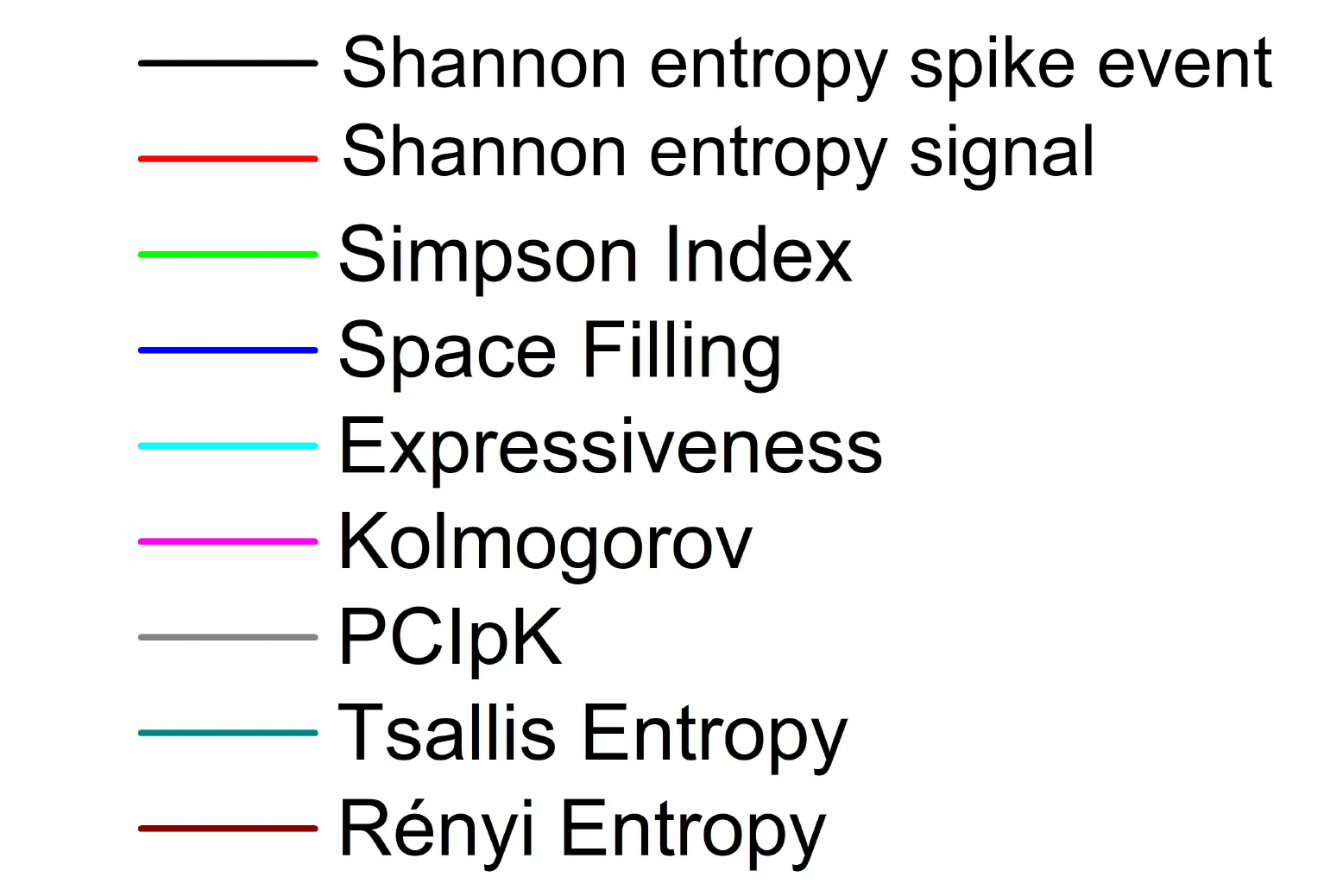}}
    \subfigure[]{\includegraphics[width=\textwidth]{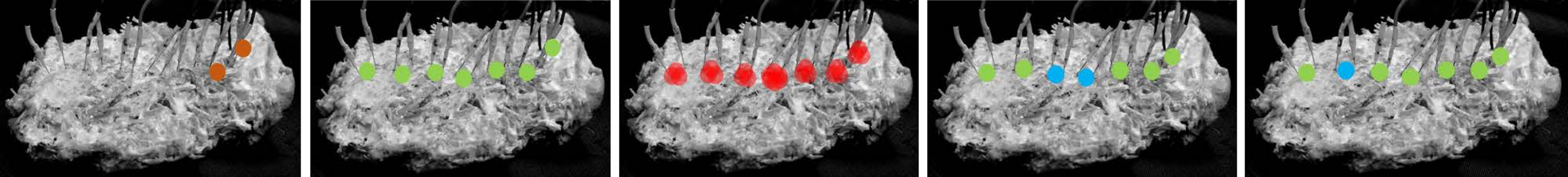}}
    \caption{(a) Evolution related to a data segment recorded 16 hours before cortisol exposure. (b) Cortisol-exposed data segment (trigger event). (c)~Evolution related to a data segment recorded 1 hour after cortisol exposure. (d) Merit figures of complexity measures performed on recorded potentials, legend. (e) Graphical representation of the living substrate with measurement electrodes and PCIpK complexity defined by a colour code (light blue $< -0.5, -0.5 <$ green $< 0$, $0 <$ orange $< 0.5$, red $> 0.5$).}
    \label{fig:multicurve}
\end{figure}

Figure~\ref{fig:multicurve} graphically summarises the qualitative description of electrical activity for all trials in a multi-curve layout that can be used to easily identify those metrics that are more influenced by stimulation with cortisol and can thus be used to track the impacts in electrical dynamics complexity. The different panels implicitly represent time, where Fig.~\ref{fig:multicurve}(a) refers to 16 hours before exposure, Fig.~\ref{fig:multicurve}(b) refers to the exposure or trigger event, and Fig.~\ref{fig:multicurve}(c) refers to 1 hour after exposure. The $x$-axis, where the channels are aligned, represents space. Those complexity metrics with no variation appear flat (for example after trigger the Shannon entropy for the signal, shown in red in the plots). Those influenced show positive or negative peaks, either to the left (Shannon entropy for the spike event, Simpson's diversity, Space-filling, Expressiveness, Kolmogorov Complexity, and PCIpK) or towards the right of the sample (Tsallis and R\'{e}nyi entropies). By superimposing complexity metrics on each channel and reconstructing the time evolution of the spiking dynamics the mycelium, we can create a graphical representation of the excitatory or inhibitory state before, during, and after cortisol exposure, demonstrating the formation of an electrical activity fingerprint that corresponds to that specific event (see Fig.~\ref{fig:multicurve}(e)).

\begin{figure}[!htbp]
    \centering
    \subfigure[]{\includegraphics[width=0.32\textwidth]{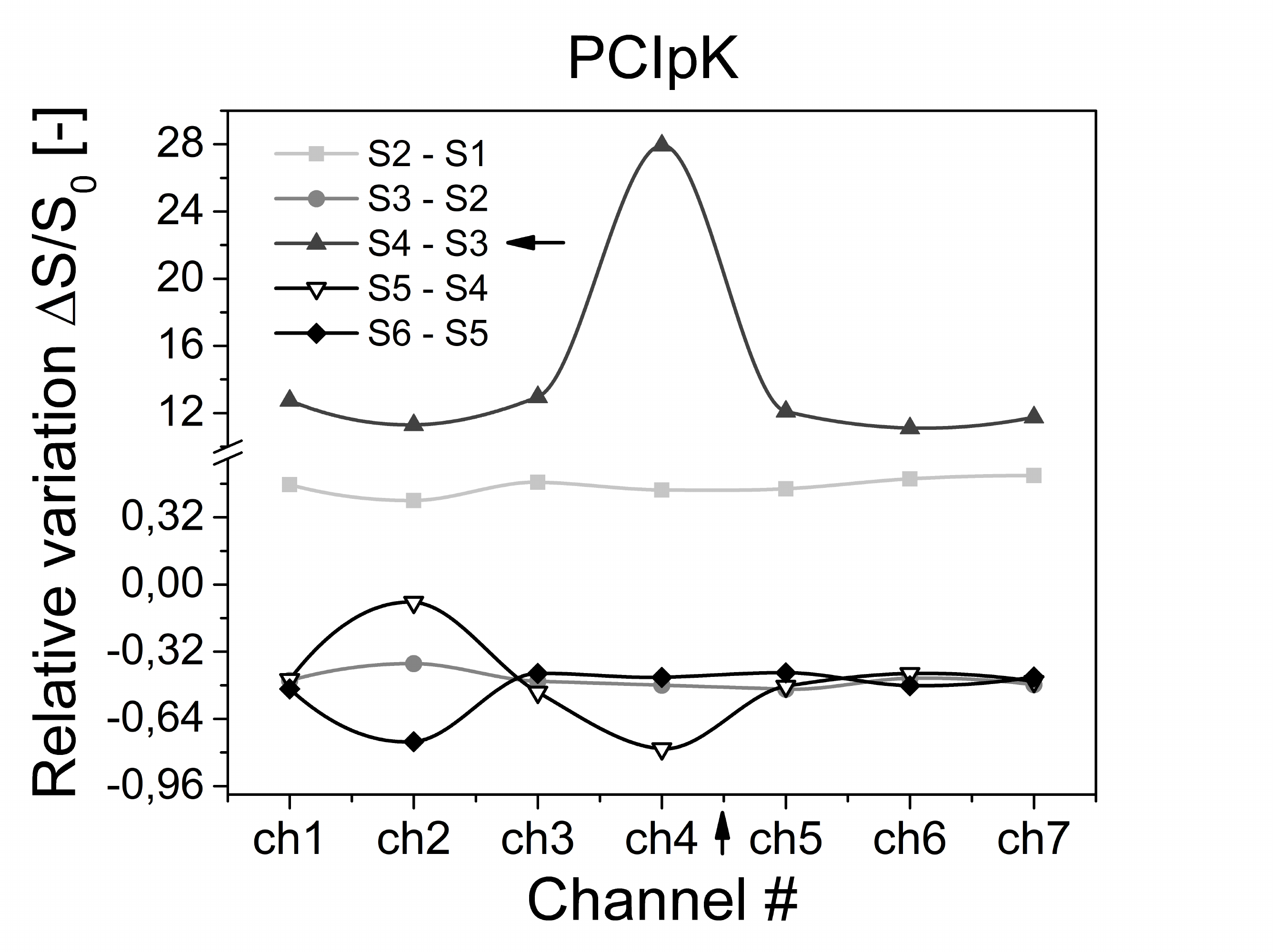}}
    \subfigure[]{\includegraphics[width=0.32\textwidth]{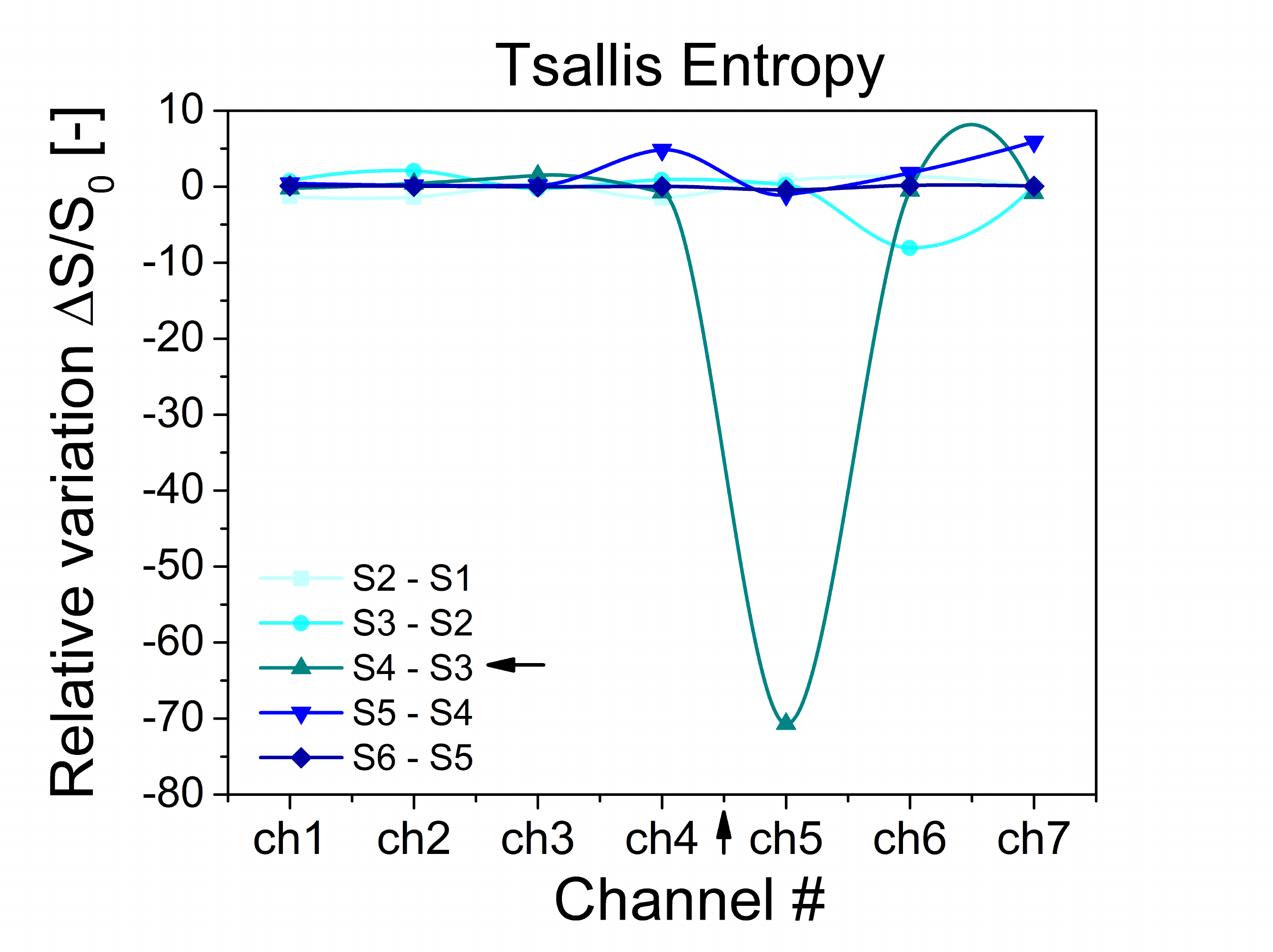}}
    \subfigure[]{\includegraphics[width=0.32\textwidth]{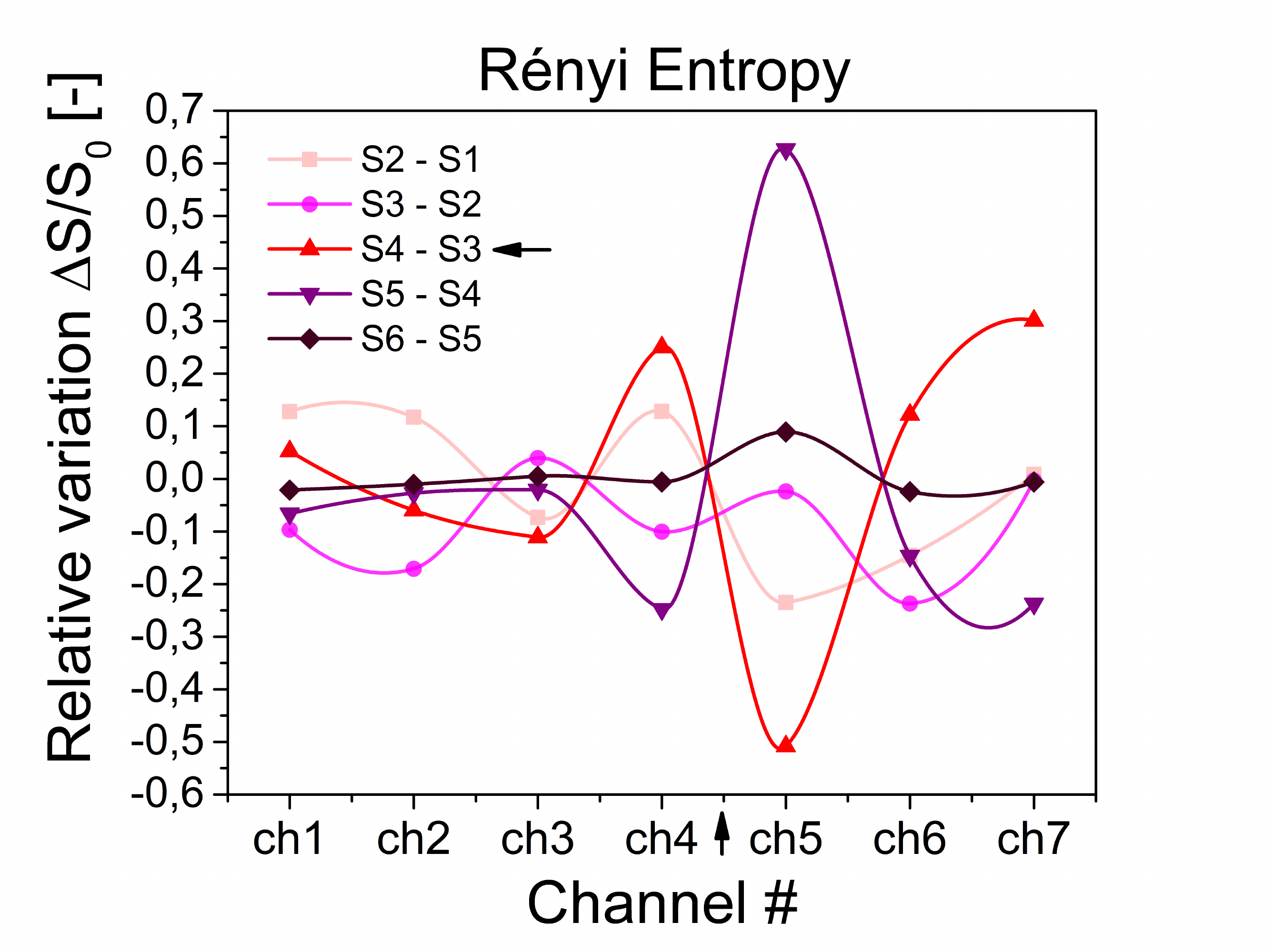}}
    \caption{(a) PCIpK is measured as a function of the measurement segments (which are time-dependent), with the arrows indicating both the spatial locus for cortisol stimulation and the temporal segment corresponding to the trigger. The time interval used to compute the relative variance is denoted by $S(n+1)-S(n)$. (b) Tsallis Entropy. (c) R\'{e}nyi Entropy.}
    \label{fig:PCIpK}
\end{figure}

We found that the PCIpK, Tsallis, and R\'{e}nyi entropies are the most relevant metrics for system analysis among all complexity measurements considered here (see Fig.~\ref{fig:PCIpK}). The PCIpK measure provides an easy inspection of the substrate's evolution analysis, including its spatio-temporal features. Figure~\ref{fig:PCIpK}(a) shows that a cortisol stimulus induces a much stronger response in the application locus. Following a stimulus, some excitation events can be seen propagating in the substrate at much lower potentials. Tsallis entropy, as shown in Fig.~\ref{fig:PCIpK}(b), is less sensitive but more accurate in tracking the evolution of the reference electrode signal. We can see from Fig.~\ref{fig:PCIpK}(c) that R\'{e}nyi entropy helps us to monitor peak evolution over time.

\subsection{Internal inspection of the fungal culture}

\begin{figure}[!htbp]
    \centering
    \includegraphics[width=1\textwidth]{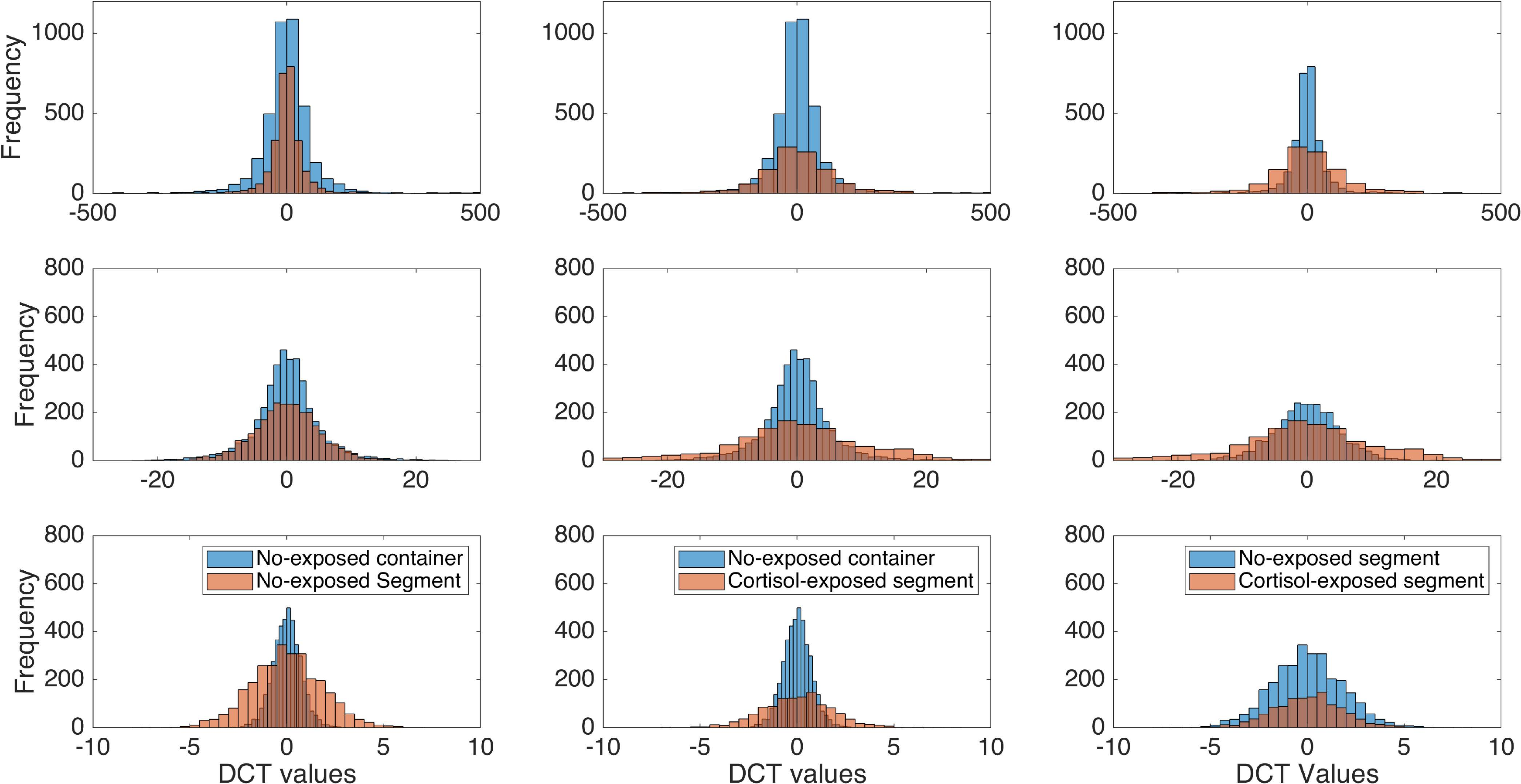}
    \caption{The first, middle, and last rows represent a comparison of the distribution of energy levels in the calculated DCT for no-exposed container, no-exposed segment, and cortisol-exposed segment, respectively.}
    \label{fig:DCTCompare}
\end{figure}

The visual appearance of the substrates colonised by the fungi did not change after exposure to cortisol. However, the impact of cortisol exposure is visible in the distribution of energy levels as a result of applying DCT to CT images, as shown in Fig.~\ref{fig:DCTCompare}. The calculated DCT values for no-exposed container, no-exposed segment, and cortisol-exposed segment were divided into three parts, including the distribution of values for high, medium, and low energies, which are shown in the first, middle, and last rows, respectively. The substrate that was not exposed to cortisol had a higher energy frequency than the cortisol-exposed segment and the no-exposed segment adjacent to the cortisol-exposed segment.

\section{Discussion} \label{sec:4}

The integrity of the fungal colony is preserved by cytoplasm flow in the mycelium network, where calcium waves~\cite{aramburu2004calcineurin} change the flow's propagation coordination. Cortisol metabolism stimulates the production of receptor activator of nuclear factor-kappa-B ligand (RANKL), a type II membrane protein that regulates bone regeneration and remodelling in mammals~\cite{chyun1984cortisol}. The activity of cells responsible for calcium resorption from bone is inhibited when RANKL is stimulated~\cite{davies1985role}. The elevated circulating cortisol levels maintain stress levels, triggering physiological changes in the body's regulatory networks. Hog1 is a stress-activated mitogen-activated protein kinase (MAPK) in fungi that is homologous to the p38 MAPK pathways in mammals~\cite{hohmann2002osmotic}. When exposed to stress conditions, Hog1 rapidly dephosphorylates and induces the appropriate cellular responses against the offending environmental stimuli~\cite{bahn2005specialization}. 
We speculate that receiving cortisol by the substrate can inhibit the flow of calcium and, in turn, reduce its physiological changes.

\section*{Acknowledgement}
This project has received funding from the European Union's Horizon 2020 research and innovation programme FET OPEN ``Challenging current thinking'' under grant agreement No 858132.

\bibliography{mybibfile}

\end{document}